\documentclass[usenatbib,useAMS]{mn2e}
\usepackage{graphicx}

\usepackage{aas_macros}
\title[]{ULTRACAM observations of SDSS J170213.26+322954.1-- an eclipsing cataclysmic variable in the period gap}

\author[]{S.\,P.\ Littlefair$^{1}$, V.\,S.\, Dhillon$^{1}$, 
T.\,R.\, Marsh$^{2}$, B.\,T.\, G\"{a}nsicke$^{2}$ \\
$^1$Dept of Physics and Astronomy, University of Sheffield, S3 7RH, UK \\
$^2$Dept of Physics, University of Warwick, Coventry, CV4 7AL, UK\\}

\date{Accepted 2006 July 4.  Received 2006 July 3; in original form 2006 April 13}

\begin{document}
\maketitle

\begin{abstract} 
We present high-speed, three-colour photometry of the eclipsing
cataclysmic variable SDSS J170213.26+322954.1 (hereafter SDSS
J1702+3229). This system has an orbital period of 2.4 hours, placing
it within the ``period gap'' for cataclysmic variables. We determine
the system parameters via a parameterized model of the eclipse fitted
to the observed light curve by $\chi^2$ minimization. We obtain a mass
ratio of $q = 0.215 \pm 0.015$ and an orbital inclination $i =
82^{\circ}.4 \pm 0^{\circ}.4$. The primary mass is $M_{\rmn{w}} =
0.94\pm0.01 M_{\sun}$. The secondary mass and radius are found to be
$M_{\rmn{r}} =0.20\pm0.01 M_{\sun}$ and $R_{\rmn{r}} = 0.243 \pm 0.013
R_{\sun}$ respectively. We find a distance to the system of $440 \pm
30$\, pc, and an effective temperature for the secondary star of
$3800\pm100$K (corresponding to a spectral type of M0$\pm$0.5V). Both
the distance and effective temperature are consistent with previous
values derived via spectroscopy of the red star.

The secondary star is significantly less massive than expected for the
orbital period, and significantly warmer than expected for its
mass. This can be explained if the secondary star is significantly
evolved: the mass and effective temperature are consistent with a
secondary star that began mass transfer with a greatly reduced central
hydrogen fraction. The nature of the secondary star in SDSS J1702+3229
supports predictions that CVs with evolved secondary stars might be
found accreting within the period gap.
\end{abstract} 


\section{Introduction}
\label{sec:introduction}
Cataclysmic variable stars (CVs) are a class of interacting binary
system undergoing mass transfer via a gas stream and accretion disc
from a Roche-lobe filling secondary to a white dwarf primary. A bright
spot is formed at the intersection of the disc and gas stream, giving
rise to an `orbital hump' in the light curve at phases $0.6-1.0$ due
to foreshortening of the bright-spot. \citet{warner95a} gives a
comprehensive review of CVs. The light curves of eclipsing CVs can be
quite complex, with the accretion disc, white dwarf and bright-spot
all being eclipsed in rapid succession. With sufficient
time-resolution, however, this eclipse structure allows the system
parameters to be determined to a high degree of accuracy \citep{wood86a}.

SDSS J1702+3229 is a deeply eclipsing CV, first discovered through the
Sloan digital sky survey \citep{szkody04}. The spectrum is highly
suggestive of a dwarf-nova type system, a fact confirmed by its recent
outburst during which it exhibited 0.3 mag superhumps (VSNET alert
8715), placing it amongst the SU UMa sub-class of Dwarf Novae.  The
system is particularly worthy of study, as its orbital period of 2.4
hours places it squarely within the period gap. Furthermore, its
deeply eclipsing nature allows accurate system parameters to be
derived. As such, SDSS J1702+3229 constitutes an excellent test of
evolutionary models for CVs, which predict secondary star masses and
radii for systems within the gap.

In this paper we present {\sc ultracam} $u'g'r'$ lightcurves of SDSS
J1702+3229, and use these lightcurves to derive the system parameters. The
observations are described in section~\ref{sec:obs}, the results are
presented in section~\ref{sec:results}, and discussed in
section~\ref{sec:disc}.

\section{Observations}
\label{sec:obs}
\begin{table*}
\begin{center}
\caption[]{Journal of observations. Observing conditions were clear
  except for 25$^{th}$ Aug 2005, when thin cirrus was present. The dead-time
  between exposures was 0.025~s for all observations. The GPS timestamping
  on each data point is accurate to 50 $\mu$s.}
\begin{tabular}{crrcccccccc}
\hline
Date & Start Phase & End Phase  & Filters & Exposure time (s) & 
Data points & Eclipses & Seeing (arcsec) & Airmass \\
\hline
2005 Aug 11 & -0.24   &  1.28   & {\em u}$^{\prime}${\em g}$^{\prime}${\em
  r}$^{\prime}$ & 1.62 & 7922 & 2 & 0.6--2.5 & 1.00--1.44 \\
2005 Aug 12 &  9.80   & 10.31   & {\em u}$^{\prime}${\em g}$^{\prime}${\em
  r}$^{\prime}$ & 1.62 & 2709 & 1 & 0.6--0.8 & 1.00--1.05 \\
2005 Aug 14 & 29.86   & 30.30   & {\em u}$^{\prime}${\em g}$^{\prime}${\em
  r}$^{\prime}$ & 1.62 & 2335 & 1 & 0.6--1.2 & 1.01--1.06 \\
2005 Aug 15 & 39.72   & 40.29   & {\em u}$^{\prime}${\em g}$^{\prime}${\em
  r}$^{\prime}$ & 1.62 & 3040 & 1 & 0.7--1.2 & 1.00--1.07 \\
2005 Aug 25 &139.65   &139.70   & {\em u}$^{\prime}${\em g}$^{\prime}${\em
  r}$^{\prime}$ & 1.62 &  372 & 0 & 0.6--1.2 & 1.02--1.03 \\
2005 Aug 25 &139.70   &140.28   & {\em u}$^{\prime}${\em g}$^{\prime}${\em
  r}$^{\prime}$ & 2.46 & 2013 & 1 & 0.7--1.2 & 1.03--1.17 \\
\hline
\end{tabular}
\label{journal}
\end{center}
\end{table*}
On nights between Aug 11$^{th}$ 2005 and Aug 25$^{th}$ 2005, SDSS
J1702+3229 was observed simultaneously in the SDSS-$u'g'r'$ colour bands
using {\sc ultracam} \citep{dhillon01} on the 4.2-m William Herschel
Telescope (WHT) on La Palma. The observations were taken between
airmasses of 1.0--1.4, in typical seeing conditions of 1.2 arcsecs,
but with a range of 0.6--2.5 arcsecs.  With the exception of the night
of Aug 25$^{th}$ 2005, when thin cirrus was present, the data were
taken in photometric conditions, albeit with heavy dust extinction
($\sim$1 mag at zenith). Six eclipses were observed in total. Data
reduction was carried out in a standard manner using the {\sc
ultracam} pipeline reduction software, as described in
\cite{feline04a}, and a nearby comparison star was used to correct the
data for transparency variations.
\section{Results}
\label{sec:results}

\subsection{Orbital Ephemeris}
\label{subsec:ephem}
\begin{table}
\begin{center}
\caption[]{Mid-eclipse timings (HJD + 2453594). The negative cycle
timings are by-eye estimates from \protect\cite{szkody04}, The
uncertainties for the values measured from our eclipses are
1.0$\times10^{-5}$.}
\begin{tabular}{rrrr}
\hline
Cycle No. & {\em u}$^{\prime}$ & {\em g}$^{\prime}$ & {\em r}$^{\prime}$\\
\hline
-7701 & \multicolumn{3}{c}{-770.341 $\pm$ 0.002} \\
-7700 & \multicolumn{3}{c}{-770.240 $\pm$ 0.002} \\
-7699 & \multicolumn{3}{c}{-770.140 $\pm$ 0.002} \\
0   &  0.39201 &  0.39204 &  0.39206 \\
1   &  0.49204 &  0.49208 &  0.49210 \\
10  &  1.39289 &  1.39285 &  1.39284 \\
30  &  3.39447 &  3.39449 &  3.39448 \\
40  &  4.39529 &  4.39528 &  4.39530 \\
140 & 14.40349 & 14.40354 & 14.40353 \\
\hline
\end{tabular}
\label{eclipse_times}
\end{center}
\end{table}
The times of white dwarf mid-ingress $T_{\rmn{wi}}$ and mid-egress
$T_{\rmn{we}}$ were determined by locating the minimum and maximum
times, respectively, of the light curve derivative.  Mid-eclipse
times, $T_{\rmn{mid}}$, were determined by assuming the white dwarf
eclipse to be symmetric around phase zero and taking
$T_{\rmn{mid}}=(T_{\rmn{we}}+T_{\rmn{wi}})/2$. These mid eclipse times
were combined with a by-eye estimate of the mid-eclipse times from
\cite{szkody04}, and a linear least-squares fit to the mid-eclipse
times was used to derive the ephemeris. The errors on the mid-eclipse
times from \cite{szkody04} were estimated by-eye, and the errors on
our mid-eclipse times were adjusted to give $\chi^{2} = 1$, giving an
ephemeris of:

\begin{displaymath}
\begin{array}{ccrcrl}
\\ HJD & = & 2453594.39209 & + & 0.10008209 & E.  \\
 & & 4 & \pm & 9 &
\end{array} 
\end{displaymath}

There was no evidence for a variation in $O-C$. This ephemeris was
used to phase our data for the analysis which follows.

\subsection{Lightcurve morphology and variations}
\label{subsec:lcurve}
\begin{figure}
\begin{center}
\includegraphics[scale=0.43,trim=0 0 0 0,clip]{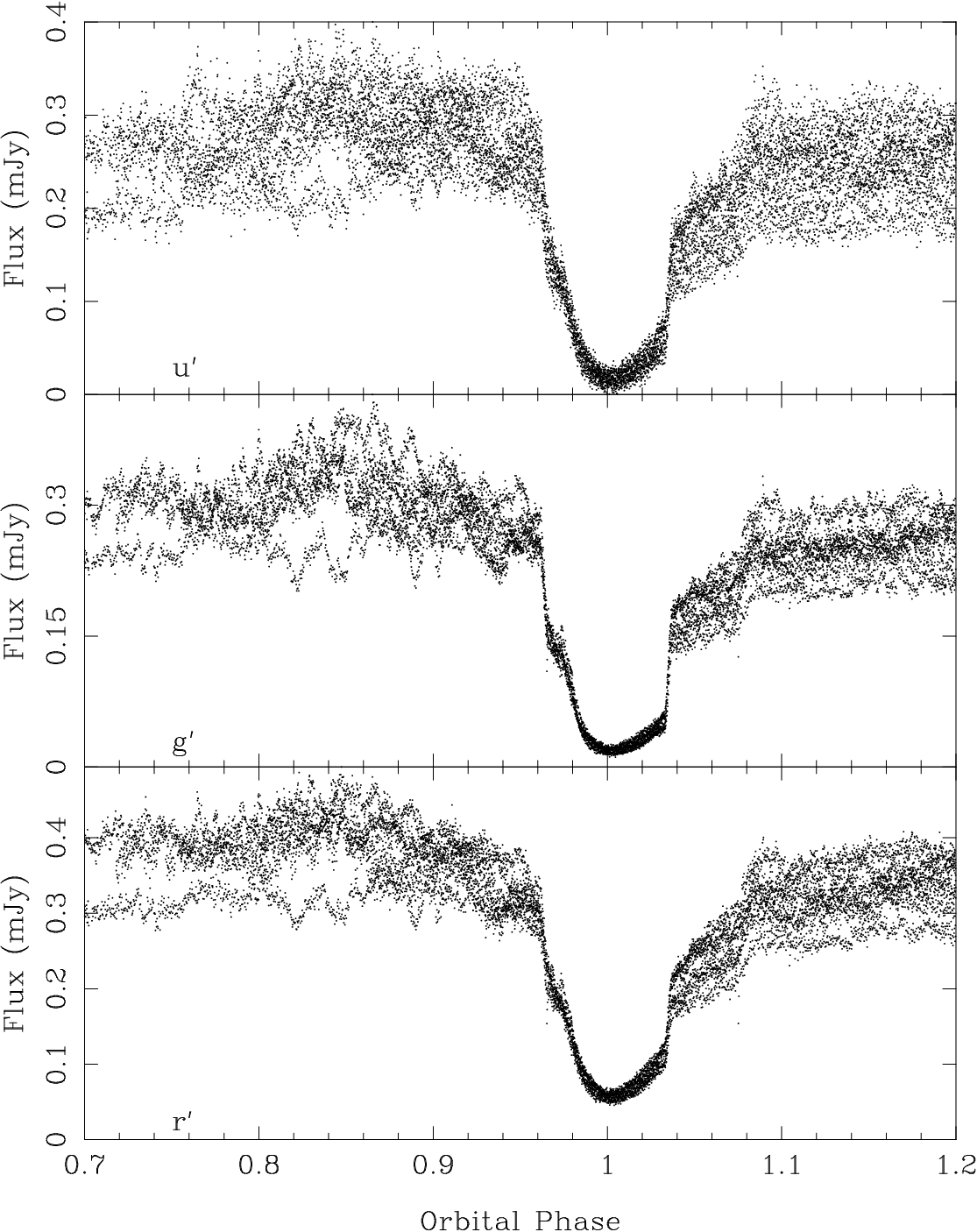} 
\caption{The phase folded $u'g'r'$ (from top to bottom) light curves
  of SDSS J1702. All six observed eclipses are shown superimposed
  to emphasise the variability between them.}
\label{fig:lcurve}
\end{center}
\end{figure}

Figure~\ref{fig:lcurve} shows the six observed eclipses of SDSS J1702,
folded on the orbital phase. The white dwarf ingress and egress
features are clearly visible, as are the ingress and egress features
of the bright spot \citep[see chapter 2.6.2 of][for an illustrated
example of an eclipse in a typical dwarf nova system]{warner95a}. The
eclipses are not total, and there is significant uneclipsed flux in
the $g'$ and $r'$ bands, suggesting the secondary star contributes
significantly at these wavelengths.There are significant drops in the
amount of variability seen in the lightcurves at phases corresponding
to the ingress of the bright spot, and the white dwarf. From this we
can conclude that there is substantial flickering from both the inner
disc and the bright spot. This finding is consistent with the results
of \cite{bruch00}, who found that the bright spot and inner disc were
the most common sites of flickering in cataclysmic variables. There is
some suggestion for variation in the shape of the orbital hump around
phase 0.8. Such variability suggests a change in large-scale structure
within the bright spot on a timescale of days. The eclipses also show
changes in the flux levels and slopes between phases 1.04--1.08, which
is strong evidence for varying disc fluxes and brightness
profiles. There is no evidence for variation in the size or duration
of white dwarf egress, as might be expected if the white dwarf is
surrounded by a boundary layer of varying thickness or flux.

\subsection{A parameterized model of the eclipse}
\label{sec:model}
To determine the system parameters we used a physical model of the
binary system to calculate eclipse light curves for each of the
various components. \cite{feline04} showed that this method gives a
more robust determination of the system parameters in the presence of
flickering than the derivative method of \cite{wood86a}. We used the
technique developed by \citet{horne94} and described in detail
therein. This model assumes that the eclipse is caused by the
secondary star, which completely fills its Roche lobe. A few changes
were necessary in order to make the model of \citet{horne94} suitable
for our data. The most important of these was the fitting of the
secondary flux, prompted by the detection of a significant amount of
flux from the secondary in the $g'$ and $r'$ bands.  We fit the model
to all the observed eclipses, which were phase-folded and binned into
groups of 7 data points.

The 10 parameters that control the shape of the light curve are as
follows:
\begin{enumerate}
\item The mass ratio, $q$.
\item The white dwarf eclipse phase full-width at half-depth, $\Delta\phi$.
\item The outer disc radius, $R_{\rmn{d}}/a$.
\item The white dwarf limb-darkening coefficient, $U_{\rmn{w}}$.
\item The white dwarf radius, $R_{\rmn{w}}/a$.
\item The bright-spot scale, $S/a$. The bright-spot is modelled as a
linear strip passing through the intersection of the gas stream and
disc. The intensity distribution along this strip is given by
$(X/S)^{2}e^{-X/S}$, where $X$ is the distance along the strip.
\item The bright-spot tilt angle, $\theta_{\rmn{B}}$,
measured relative to the line joining the white dwarf and the
secondary star. This allows adjustment of the phase of the orbital hump.
\item The fraction of bright-spot light which is isotropic, $f_{iso}$.
\item The disc exponent, $b$, describing the power law of the radial
intensity distribution of the disc.
\item A phase offset, $\phi_{0}$.
\end{enumerate}

The {\sc amoeba} algorithm (downhill simplex; \citealt{press86}) was
used to adjust selected parameters to find the best fit. A linear
regression was used to scale the four light curves (for the white
dwarf, bright-spot, accretion disc and secondary) to fit the observed
light curves in each passband. The data were not good enough to
determine the white dwarf limb-darkening coefficient, $U_{\rmn{w}}$,
accurately. To find an appropriate limb-darkening coefficient, we
obtained an estimate of the effective temperature and mass of the
white dwarf from a first iteration of the method below, and assuming a
limb-darkening coefficient of 0.5. The mass and effective temperature
was then used in conjunction with the stellar atmosphere code of
\cite{gaensicke95} to generate angle-dependent white dwarf model
spectra. To convert the spectra to observed fluxes the model spectra
were folded through passbands corresponding to the instrumental
response in each filter; the effects of the SDSS filter set, the {\sc
ultracam} CCD responses and the dichroics used in the instrumental
optics were taken into account. These fluxes were then fit as a
function of the limb position in order to derive limb-darkening
parameters for each band. The values adopted are given in
table~\ref{parameters_lfit}. A second iteration using these values for
the limb-darkening parameter gave the final values for each parameter.

In order to estimate the errors on each parameter once the best fit
had been found, we perturbed one parameter from its best fit value by
an arbitrary amount (initially 5 per~cent) and re-optimised the rest
of them (holding the parameter of interest, and any others originally
kept constant, fixed). We then used a bisection method to determine
the perturbation necessary to increase $\chi^{2}$ by $1$, i.e.\
$\chi^{2}-\chi_{\rmn{min}}^{2}=\Delta\chi^{2}=1$. The difference
between the perturbed and best-fit values of the parameter gave the
relevant $1\sigma$ error \citep*{lampton76}.

\begin{figure}
\begin{center}
\includegraphics[scale=0.35,angle=-90,trim=0 30 20 0,clip]{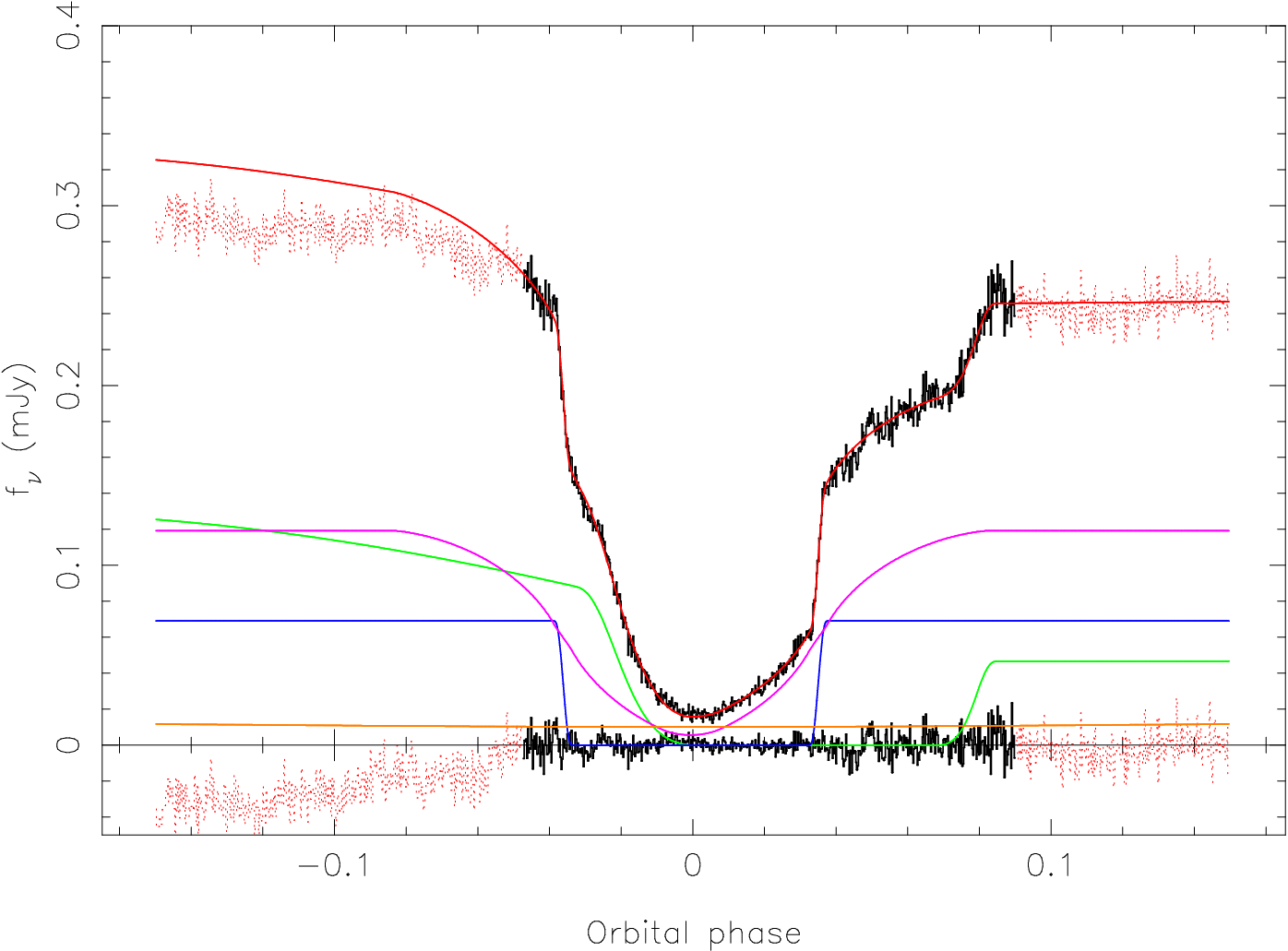} 
\includegraphics[scale=0.35,angle=-90,trim=0 30 20 0,clip]{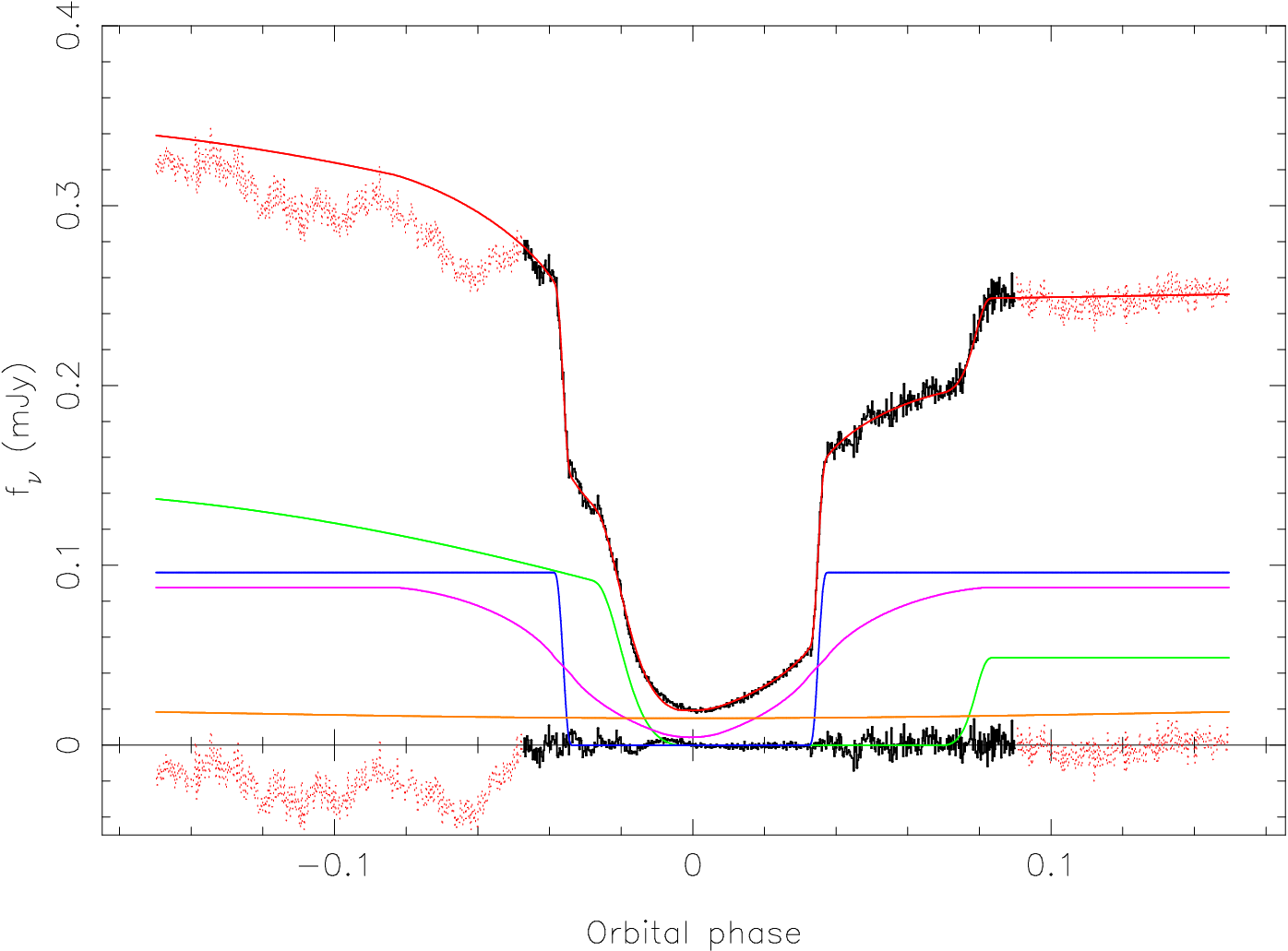} 
\includegraphics[scale=0.35,angle=-90,trim=0 30 -20 0]{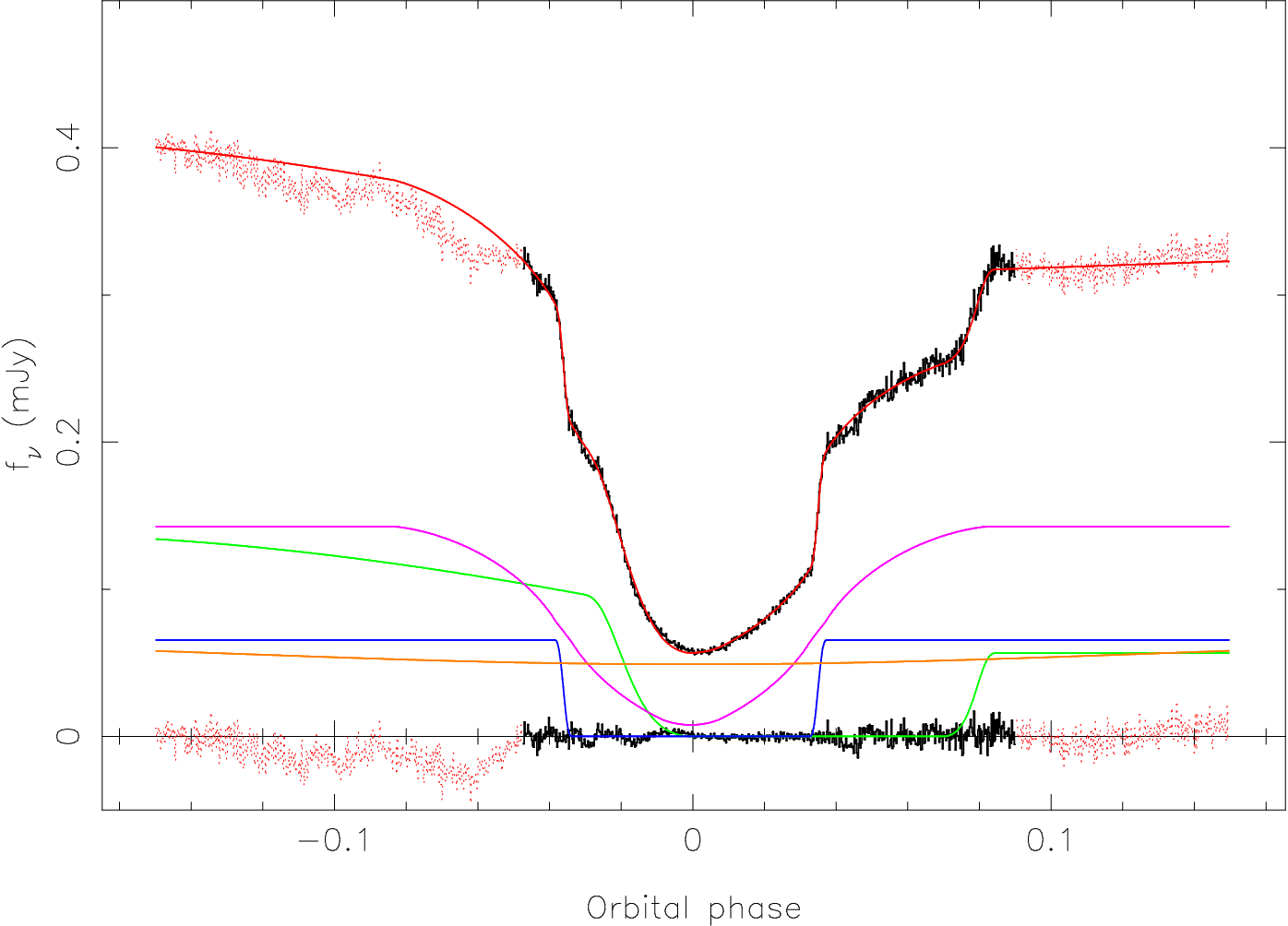} 
\caption{The phased folded $u'g'r'$ (from top to bottom) light curves
  of SDSS J1702, fitted separately using the model described in
  section~\ref{sec:model}. The data (black) are shown with the fit
  (red) overlaid and the residuals plotted below (black). Below are
  the separate light curves of the white dwarf (blue), bright spot
  (green), accretion disc (purple) and the secondary star
  (orange). Data points neglected in the fit are shown in red (light grey). }
\label{fig:model}
\end{center}
\end{figure}

\begin{table*}
\begin{center}
\caption[]{Parameters fitted using a modified version of the model of
\citet{horne94}. The fluxes of each component are also shown. Prior to
fitting, the data were phase-folded and binned by seven data points.
Note that the orbital inclination $i$ is not a fit parameter but is
calculated using $q$ and $\Delta\phi$.}
\begin{tabular}{lccc}
\hline
Band & {\em u}$^{\prime}$ & {\em g}$^{\prime}$ & {\em r}$^{\prime}$ \\
\hline
Inclination $i$ & $82\fdg0\pm0\fdg1$ & $82\fdg6\pm0\fdg1$ & $82\fdg6\pm0\fdg1$ \\
Mass ratio $q$ & $0.226\pm0.002$ & $0.210\pm0.001$ & $0.209\pm0.001$ \\
Eclipse phase width $\Delta\phi$& $0.0713\pm0.0001$ & $0.07126\pm0.00003$ & $0.0711\pm0.0001$ \\
Outer disc radius $R_{\rmn{d}}/a$ & $0.290\pm0.003$ & $0.288\pm0.001$ & $0.293\pm0.001$\\
White dwarf limb darkening $U_{\rmn{w}}$ & $0.344\pm0.008$ & $0.267\pm0.008$ & $0.227\pm0.003$ \\
White dwarf radius $R_{\rmn{w}}/a$ & $0.0095\pm0.0007$ & $0.0096\pm0.0001$ & $0.0088\pm0.0006$ \\
bright-spot  scale $S/a$ & $0.043\pm0.001$ & $0.0338\pm0.0004$ & $0.040\pm0.001$\\
bright-spot  orientation $\theta_{\rmn{B}}$ & $161\fdg4\pm0\fdg6$ & $162\fdg4\pm0\fdg3$ & $161\fdg4\pm0\fdg7$ \\
Isotropic flux fraction $f_{iso}$ & $0.36\pm0.01$ & $0.343\pm0.003$ & $0.410\pm0.003$ \\  
Disc exponent $b$ & $-1.1\pm0.1$ & $-1.1\pm0.1$ & $-1.0\pm0.1$ \\ 
Phase offset $\phi_{0}$ & $57\pm3\times10^{-5}$ & $60\pm2\times10^{-5}$ & $57\pm2\times10^{-5}$ \\
$\chi^{2}$ of fit & 866 & 4634 & 3555 \\
Number of datapoints $\nu$ & 1268 & 1268 & 1268 \\
\hline
Flux (mJy) \\
\hspace{0.1cm} White dwarf    & $0.0690\pm0.0010$ & $0.0960\pm0.0003$ & $0.0654\pm0.0004$ \\
\hspace{0.1cm} Accretion disc & $0.1191\pm0.0010$ & $0.0877\pm0.0004$ & $0.1427\pm0.0005$ \\
\hspace{0.1cm} Secondary      & $0.0126\pm0.0004$ & $0.0203\pm0.0001$ & $0.0629\pm0.0002$ \\
\hspace{0.1cm} bright-spot     & $0.1294\pm0.0007$ & $0.1418\pm0.0002$ & $0.1379\pm0.0003$ \\
\hline
\end{tabular}
\label{parameters_lfit}
\end{center}
\end{table*}

The results of the model fitting are given in
table~\ref{parameters_lfit} and shown in figure~\ref{fig:model}.  It
is apparent that the model is a very good fit to the data during
eclipse. In particular, the ingress and egress features are well
modelled.  Further confidence in the results of the fitting is
provided by the consistency in fitted parameters between wavebands
(for example, the white dwarf radius and eclipse phase
width). However, the model is a poor fit to the data prior to
eclipse. Given the large changes seen in this part of the lightcurve
between different eclipses (see figure~\ref{fig:lcurve}), this is
unsurprising, but it may also reflect the simple nature of the disc
and bright spot in our model.  To limit the effects that the poor fit
to pre-eclipse data may have on the resulting system parameters we
excluded the regions shown in red (light grey) in
figure~\ref{fig:model} from the fit.  The exclusion of these regions
from the fit can have a small effect on the derived value of the mass
ratio, $q$. In our model, constraints on the mass ratio arise from the
position of bright spot ingress and egress, but also from the height
of the bright-spot hump seen before eclipse, and the shape of the
lightcurve at mid eclipse. The net effect is that the values for $q$
derived from each band are marginally inconsistent, with a difference
of around $4\sigma$. This inconsistency in $q$ produces a
corresponding uncertainty in $i$, which depends upon $q$ and
$\Delta\phi$. In the analysis that follows we assume that the scatter
in $q$ between SDSS bands is indicative of the uncertainty in this
parameter, and adopt values of $q=0.215\pm 0.01$ and
$i=82\fdg4\pm0\fdg4$, as shown in table~\ref{parameters}.

We calculated the remaining system parameters following the method
described in \cite{feline04}. In brief, the white dwarf fluxes in
table~\ref{parameters_lfit} are fitted using $\chi^2$ minimisation to
white dwarf colours from the model atmospheres of
\cite{bergeron95}, converted to the SDSS system using the
transformations of \cite{smith02}; this provides an estimate of the
white dwarf temperature. The white dwarf mass can then be derived
using a white dwarf mass--radius relation, and the remaining system
parameters are derived from the white dwarf mass and the results of
the model fits. We used the Nauenberg mass-radius relation
\citep{nauenberg72}. Because the Nauenberg relation applies to a cold,
non-rotating white dwarf we applied a correction to the mass-radius
relation, appropriate for the white dwarf temperature derived
above. The radius of the white dwarf at $10000$~K is about 5 per~cent
larger than for a cold ($0$~K) white dwarf \citep{koester86}. To
correct from $10000$~K to the appropriate temperature, the white dwarf
cooling curves of \citet{wood95} for $M_{\rmn{w}}/M_{\sun}=1.0$, the
approximate mass given by the Nauenberg relation, were used. This
gave a radial correction factor of $7.0$ per~cent.  A distance to the
system was derived by comparing the white dwarf fluxes in
table~\ref{parameters_lfit}, and the predicted fluxes from
\cite{bergeron95}; the uncertainty in the distance is dominated by the
uncertainty in the white dwarf temperature. The derived distance of
$440 \pm 30$\,pc is consistent with the lower end of possible
distances derived by \cite{szkody04} of 460--650\,pc. The latter
estimate was obtained by assuming the secondary star contributes 100\%
of the red light in the system, and so we might expect the true
distance to lie towards the lower end of the range.

In addition, we attempted to derive an effective temperature for the
secondary star, by fitting the secondary star fluxes to the colours of
main-sequence stars. This method should be treated with caution, as
the measured flux from the secondary star depends upon assumptions
about the disc emission and bright-spot lightcurve. Indeed, comparison
of the secondary star colours with the colours of main-sequence stars
in the SDSS photometric system \citep{fukugita96} showed that the
$u'-g'$ colour was exceedingly blue for a main sequence star, given
the $g'-r'$ colour. Working under the assumption that the $g'-r'$
colour is much less likely to be contaminated by disc and bright spot
emission, we used $\chi^2$ minimisation to fit the $g'-r'$ colour to
the colours of main-sequence stars as catalogued by \cite{kenyon95}.
The effective temperature we derive agrees  well with the
spectral type of M1.5$\pm$1 derived from optical spectroscopy \citep{szkody04},
giving us confidence that the temperature is correct. The final
adopted system parameters are shown in table~\ref{parameters}.

\begin{table}
\begin{center}
\caption[]{System parameters of SDSS J1702+3229 derived using the Nauenberg
  mass--radius relation corrected to the appropriate
  $T_{\rmn{w}}$. $R_{\rmn{r}}$ is the volume radius of the secondary's
  Roche lobe \citep{eggleton83}. Except for the mass ratio and the
  inclination, which are discussed in section~\ref{sec:model}, the
  weighted means of the appropriate values from
  Table~\ref{parameters_lfit} are used for the system parameters. }
\begin{tabular}{lcccc}
\hline
\hline
Inclination $i$ & $82\fdg4 \pm 0\fdg4$\\
Mass ratio $q=M_{\rmn{r}}/M_{\rmn{w}}$ & $0.215 \pm 0.015$ \\
White dwarf mass $M_{\rmn{w}}/M_{\sun}$ & $0.94 \pm 0.01$ \\
Secondary mass $M_{\rmn{r}}/M_{\sun}$ & $0.20 \pm 0.01$ \\
White dwarf radius $R_{\rmn{w}}/R_{\sun}$ & $0.0091 \pm 0.0001$ \\
Secondary radius $R_{\rmn{r}}/R_{\sun}$ & $0.243 \pm 0.013$ \\
Separation $a/R_{\sun}$ & $0.948 \pm 0.005$ \\
White dwarf radial velocity $K_{\rmn{w}}/\rmn{km\;s^{-1}}$ & $84 \pm 6$ \\
Secondary radial velocity $K_{\rmn{r}}/\rmn{km\;s^{-1}}$ & $391 \pm 1.0$ \\
Outer disc radius $R_{\rmn{d}}/a$ & $0.290 \pm 0.001$ \\
White dwarf temperature $T_{\rmn{w}}/\rm{K}$ & $17000 \pm 500$ \\
Secondary star temperature $T_{\rmn{r}}/\rm{K}$ & $3800 \pm 100$ \\
Distance $d/\rmn{pc}$ & $440 \pm 30$ \\
\hline
\end{tabular}
\label{parameters}
\end{center}
\end{table}

\section{Discussion}
\label{sec:disc}

\subsection{Superhump period excess and mass ratio}
\label{subsec:superhumps}
On the 3$^{rd}$ Oct 2005, SDSS J1702+3229 entered its first recorded
outburst (VSNET Alert 8709); On Oct 7$^{th}$, 0.3 mag superhumps were
observed, with a superhump period $P_{sh}$ of 0.1056 days (VSNET Alert
8715). The detection of superhumps in this object is significant in
the context of the well-established relationship between mass ratio
and superhump period excess, as very few systems are available to
calibrate the relationship at high mass ratios \citep{patterson05}.

From the reported superhump period and the orbital period reported in
this paper we calculate a superhump period excess, $\epsilon = 0.0551
\pm 0.0005$. This is entirely consistent with the $\epsilon$--$q$
relationship established by \cite{patterson05} which predicts
$\epsilon = 0.052 \pm 0.005$ for the mass ratio of SDSS J1702+3229.

\subsection{Evolutionary status of the secondary star}
\label{subsec:evolved}

Several independent theoretical studies predict the existence of a
population of CVs with substantially evolved secondary stars
\citep[e.g][]{baraffe00,andronov04,schenker02,podsiadlowski03}. Under
the disrupted magnetic braking model of CV evolution, the period gap
is caused by a cessation of magnetic braking which occurs when the
secondary star becomes fully convective. The small central hydrogen
abundance and higher central temperature of evolved stars imply lower
radiative opacities; this naturally favours radiative
transport. Sufficiently evolved secondary stars do not, therefore,
become fully convective until they have evolved to shorter periods
than the upper edge of the period gap. For sufficiently evolved
secondary stars, the system can continue accreting throughout the gap
\citep{baraffe00}. We might therefore expect that a sizeable
proportion of CVs with periods inside the period gap would possess
evolved secondary stars.

Evidence that the secondary star in SDSS J1702+3229 has undergone
significant nuclear evolution comes from two independent lines of
argument; the secondary star is both too warm and insufficiently
massive to be an ordinary star, given its orbital period. In this
paper, we find an effective temperature for the secondary star of
$3800\pm100$\,K, which agrees well with the reported spectral type of
M1.5$\pm$1, or $3600\pm200$\,K \citep{szkody04}, derived from the TiO
band spectral index derived by \cite{reid95}.  An effective
temperature of 3800\,K is too warm for a non-evolved CV secondary star
at an orbital period of 2.4 hours \citep{baraffe00}. At periods of
around 3 hours, a typical spectral type is approximately M4 (c.\,f.\,
U Gem and IP Peg), whereas a spectral type of M1 is more typical of
periods near 5 hours \cite[see][for examples]{sad98}.  On the other
hand, the early spectral type {\em is} consistent with a secondary
star which has undergone significant nuclear evolution. From figure~3
of \cite{baraffe00}, it is clear that the effective temperature of the
secondary star in SDSS J1702+3229 is consistent with models of a
secondary star that began mass transfer with a greatly reduced central
hydrogen fraction.

The mass of the secondary star at a given orbital period is also
indicative of the evolutionary state; the secondary mass is, to first
order, a simple function of the period and the mass-radius
relationship \citep[see][for example]{howell01a}. The consequence of
this is that an evolved secondary star will be under-massive for a
given period, compared with the expected mass of a main-sequence
secondary star. In fact, this is precisely what we observe; adopting
the main-sequence mass-radius relationship of \cite{chabrier97}, we
find the secondary star mass at a period of 2.4 hours should be
$0.25M_{\sun}$. In this paper we find a secondary star mass of
$M_{\rmn{r}} =0.20\pm0.01 M_{\sun}$, which is significantly lower than
expected for a main sequence star.

It is worth spending some time speculating upon the origin and
subsequent evolution of SDSS J1702+3229.  Clearly the secondary star
has undergone significant nuclear evolution; the requirement for this
to happen within a Hubble time implies that the initial mass of the
secondary should be greater than 0.8$M_{\sun}$. What happened to the
system after contact depends upon the secondary mass and upon the
evolutionary state of the secondary at contact; for a primary mass of
0.94$M_{\sun}$, an initial secondary mass of 1.2$M_{\sun}$ or less
implies that mass transfer {\em from a main sequence star} will be
both thermally and dynamically stable \citep{politano96}. However, if
the secondary star had already left the main sequence before contact,
or was more massive than 1.2$M_{\sun}$, then the system would have
undergone a phase of thermal timescale mass transfer. During this
phase the orbit would shrink rapidly. Once mass transfer regained
stability, the system would become recognisable as a CV at a much
lower period. This channel raises the possibility that SDSS J1702+3229
might have emerged from thermal timescale mass transfer within the
period gap. The accurate values of secondary mass, radius and
effective temperature presented here should allow detailed modelling
to recover the evolutionary past of SDSS J1702+3229, however this is
beyond the scope of this paper.  The future evolution of SDSS
J1702+3229 depends sensitively on the evolutionary state of the
secondary star. However, comparison with the models of
\cite{podsiadlowski03} suggests that the system will evolve below the
observed ``period-minimum'', and may even be a progenitor of an AM CVn
system.

\subsection{Evolved secondary stars in CVs}
A small number of CV systems now show very strong evidence for the
presence of an evolved secondary star; both QZ Ser and EI Psc have
unusually hot secondaries for their orbital period
\citep{thorst02a,thorst02b}, and the secondary star in EI Psc has a
very large N/C abundance \citep{gaensicke03}. Indeed, the period of EI
Psc is substantially below the period minimum, which is in principle
only possible for an evolved secondary star. Given the difficulty in
obtaining information about the secondary stars in CVs, it is at least
plausible that systems with evolved secondary stars constitute a
significant fraction of the CV population, although it is not yet
possible to state whether the fraction is as high as 10\%, as
suggested by \cite{podsiadlowski03}.

The existence of evolved secondaries amongst the cataclysmic variable
population, and the fact that such systems may not show a period gap
provides a natural explanation for the non-magnetic CVs found within
the gap. It is tempting to speculate that most non-magnetic CVs within
the period gap are explained by systems with evolved secondary
stars. Given that the first such system which allows accurate
determination of system parameters has an evolved secondary there may
be some support for this hypothesis. Furthermore, such a hypothesis
might explain the large fraction of systems within the period gap
which exhibit superhumps \citep{katysheva03}; superhumping systems
must have mass ratios below a critical value of approximately 0.3, and
the undermassive secondary stars in evolved systems would increase the
proportion of systems with mass ratios which are unstable to
superhumps.

We conclude that the secondary star in SDSS J1702+3229 shows evidence
for significant nuclear evolution. The existence of a CV with an
evolved secondary within the period gap supports predictions
\citep[e.g.][]{baraffe00} that CVs with evolved secondaries can
continue accreting inside the period gap, and in some cases might show
no period gap at all.

\section*{\sc Acknowledgements}
TRM acknowledges the support of a PPARC Senior Research
Fellowship. BTG acknowledges the support of a PPARC Advanced
Fellowship.  ULTRACAM and SPL are supported by PPARC grants
PP/D002370/1 and PPA/G/S/2003/00058, respectively. This research has
made use of NASA's Astrophysics Data System Bibliographic
Services. Based on observations made with the William Herschel
Telescope operated on the island of La Palma by the Isaac Newton Group
in the Spanish Observatorio del Roque de los Muchachos of the
Instituto de Astrofisica de Canarias.

\bibliographystyle{mn2e}
\bibliography{abbrev,refs,refs2}

\end{document}